\documentclass[12pt,preprint]{aastex}

\shortauthors{Wang et al.}
\begin{document}

\title{Relativistic outflow in CXO CDFS J033260.0-274748}
\author{J. X. Wang\altaffilmark{1,2},
T. G. Wang\altaffilmark{1},
P. Tozzi\altaffilmark{3},
R. Giacconi\altaffilmark{2},
G. Hasinger\altaffilmark{4},
L. Kewley\altaffilmark{5},
V. Mainieri\altaffilmark{4},
M. Nonino\altaffilmark{3},
C. Norman\altaffilmark{2,6},
A. Streblyanska\altaffilmark{4},
G. Szokoly\altaffilmark{4},
T. Yaqoob\altaffilmark{2,7},
and
A. Zirm\altaffilmark{8}
}
\altaffiltext{1}{Center for Astrophysics, University of Science and Technology of China, Hefei, Anhui 230026, P. R. China; jxw@ustc.edu.cn.}
\altaffiltext{2}{Dept. of Physics and Astronomy, The Johns Hopkins University,
Baltimore, MD 21218}
\altaffiltext{3}{INAF Osservatorio Astronomico, Via G. Tiepolo 11, 34131 Trieste, Italy}
\altaffiltext{4}{Max-Planck-Institut f\"{u}r extraterrestrische Physik, Postfach 1312, D-85741 Garching, Germany}
\altaffiltext{5}{Hubble Fellow, Institute for Astronomy, University of Hawaii, 2680 Woodlawn Drive, Manoa, HI 96822}
\altaffiltext{6}{Space Telescope Science Institute, 3700 San Martin Drive, Baltimore, MD 21218}
\altaffiltext{7}{Laboratory for High Energy Astrophysics, NASA/GSFC, code 662, Greenbelt, MD 20771}
\altaffiltext{8}{Department of Astronomy, Leiden Observatory, P. O. Box 9513, 2300 RA, Leiden, The Netherlands}
\vspace{0.1cm}

\begin{abstract}
In this letter we report the detection of a strong and extremely
blueshifted X--ray absorption feature in the 1 Ms $Chandra$ spectrum
of CXO CDFS J033260.0-274748, a quasar at $z$ = 2.579 with $L_{2-10
keV} \sim 4 \times 10^{44}$ ergs s$^{-1}$.  The broad absorption
feature at $\sim 6.3$ keV in the observed frame can be fitted either
as an absorption edge at 20.9 keV or as a broad absorption line at
22.2 keV rest frame.  The absorber has to be extremely ionized with
an ionization parameter $\xi \sim 10^4$, and a high column density $N_H >
5 \times 10^{23}$ cm$^{-2}$.  We reject the possibility of a
statistical or instrumental artifact.  The most likely interpretation
is an extremely blueshifted broad absorption line or absorption edge,
due to H or He--like iron in a relativistic jet-like outflow with bulk velocity
of $\sim 0.7 - 0.8 c$.  Similar relativistic outflows
have been reported in the X--ray spectra of several other
AGNs in the past few years.
\end{abstract}

\section{Introduction}
The 1 Ms exposure of $Chandra$ Deep Field South (CDFS, e.g. Giacconi
et al.  2002; Rosati et al. 2002), along with the 2 Ms exposure of
$Chandra$ Deep Field North (CDF--N, e.g.  Brandt et al. 2001;
Alexander et al. 2003) are the deepest X--ray image ever taken.  With
these two deep surveys, the cosmic X--ray background discovered by
Giacconi et al. (1962), the origin of which has been a major goal of
X--ray astronomy for almost four decades, is now almost completely
resolved into individual sources. Most of these X--ray sources are
extragalactic, harboring supermassive black holes. The major goal now
is studying the properties of these X--ray sources and understanding
their physical nature.

Most recent $Chandra$ and $XMM$ observations have detected relativistic 
outflows with bulk velocity of $v_{\rm out} \sim$ $0.6 - 0.7 c$ in several active galactic nuclei 
(Yaqoob et al. 1998, 1999; Wang et al. 2003), 
revealed by extremely blueshifted emission lines due to iron or other
elements in the X-ray spectra.  
In this letter, we report the discovery of a
strong and blueshifted X--ray absorption feature (suggesting an even higher
$v_{\rm out} \sim$ $0.7 - 0.8 c$) in one of the CDF--S sources, CXO CDFS J033260.0-274748
(hereafter CDFS11).

\section{The Data and X--ray spectral fitting}
The 1~Ms $Chandra$ exposure on CDF-S was
composed of eleven individual ACIS observations obtained from
October 1999 to December 2000.
Giacconi et al. (2002) presented the detailed X--ray data reduction
and the final X--ray catalog (see also Alexander et
al. 2003). 
The source CDFS11, 
$\sim$ 7$\arcmin$ from the center of the field, was observed in
all the eleven exposures.
Its X--ray radial intensity profile is consistent with that of a
point-source (Giacconi et al. 2002), and no other X--ray source was
detected within $45\arcsec$.  The optical counterpart of CDFS11
($R_{vega} = 21.8$) was selected using the deep optical image
($R_{Vega} <26$), obtained with the FORS1 camera on the ANTU telescope
(UT-1 at VLT). Giacconi et al. (2002) presented the $R$--band image
cut-out, overplotted with X--ray flux contours (see their Fig. 13r).
We can clearly see a single point--like optical counterpart located
right at the center of the X--ray contours, with no nearby optical
source within $7\arcsec$.  The source was firmly classified as a
quasar at $z = 2.579$ by the follow--up spectroscopy observations
(see Fig. 6 of Szokoly et al. 2004). 
The X--ray to optical flux ratio log($f_{2-10 keV}/f_R$) is 0.3,
typical for AGNs (see, e.g. Hornschemeier et al. 2001).  A weak,
flat--spectrum radio counterpart has also been detected with the VLA
with flux densities of $36 \pm 10$ microJy at 6cm, and $33 \pm 11$ at
22cm (Kellermann et al. in preparation), making it radio quiet with 
radio--to--optical ratio $f_{1.4GHz}/f_R\sim$ 6.

\begin{figure}
\plotone{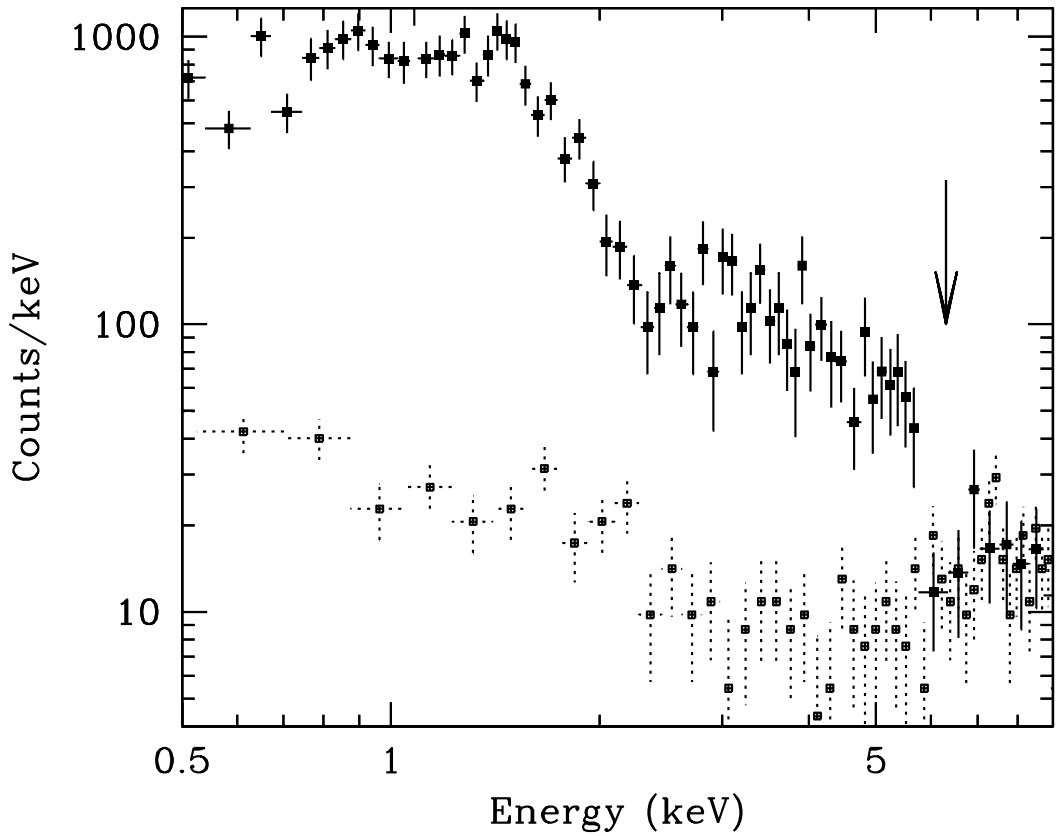}
\caption{
The summed (source plus background) X--ray spectrum of CDFS11
and the expected background (points with dotted error bars).
The spectra were rebinned for display purpose.
}
\label{spec}
\end{figure}

We extract the $Chandra$ ACIS--I X--ray spectrum of CDFS11 from a
circle with radius of 7$\arcsec$, which is the 95\% encircled-energy
radius of the ACIS point--spread function at the source position. The
local background was extracted from an annulus with outer radius of
$19\arcsec$ and inner radius of $9\arcsec$.  In Fig. \ref{spec} we
present the summed spectrum (source plus background) and the
background evaluated in the outer annulus.
The source is fairly bright in X--ray with $\sim 1040$ net X--ray
counts in the $0.5 - 2.0$ keV band and $\sim 350$ in the $2.0 - 9.0$ keV
band, allowing us to perform X--ray spectroscopy.  During the fit, we
use the C statistics (Cash 1979; Nousek \& Shue 1989), which has better
performance with respect to the $\chi^2$ analysis, particularly for
spectra with low detected counts.
We generate the X--ray telescope response and ACIS--I instrument
response for each single $Chandra$ observation, and sum the response
files weighting them for the corresponding exposure times.  The final
time--weighted response files were used for spectral analysis.  We
utilize XSPEC v11.2 to perform the spectral fitting.  All the spectral
fitting was done in the energy band $0.5 - 9.0$ keV and all the
statistical errors quoted in this paper are at the 90\% confidence
level for one interesting parameter.

\begin{figure}
\plotone{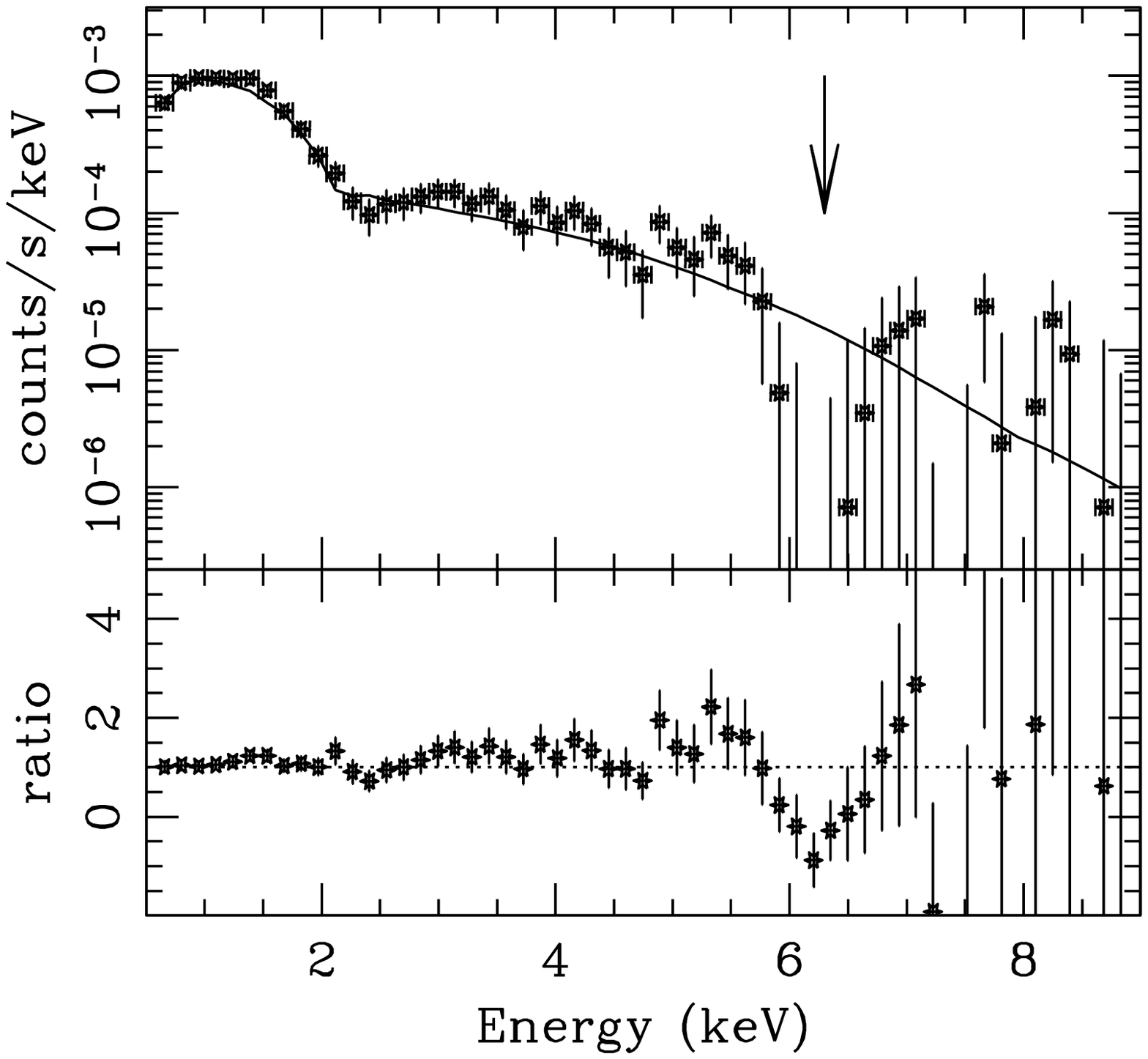}
\caption{The spectral data (rebinned for display purpose), best-fit continuum models and the ratios of
data to model for CDFS11.
}
\label{ratio}
\end{figure}

We first fitted the spectrum with a simple power--law plus a neutral
absorber in the source frame. A Galactic neutral hydrogen absorption
column of $8 \times 10^{19} \ \rm cm^{-2}$ (Dickey \& Lockman 1990)
was also included.  The results are presented in Table 1. The spectrum
was well fitted by a power--law ($\Gamma = 1.7^{+0.1}_{-0.1}$) with
weak absorption ($N_H = 0.1^{+1.0}_{-0.1} \times 10^{22}$ cm$^{-2}$).
The intrinsic, rest--frame $2.0 - 10.0$ keV luminosity is $4.3 \times
10^{44}$ ergs s$^{-1}$ ($H_0 = 70.0$ km s$^{-1}$Mpc$^{-1}$, $\Omega_m$
= 0.27, $\Omega_\Lambda$ = 0.73).  The best--fit continuum model and
the ratio of data to model is shown in Fig. \ref{ratio}.
Both in Fig \ref{spec} and \ref{ratio}, we can see a strong absorption
feature at $\sim$ 6.3 keV in the observed frame. The absorption
appears to be optically thick, since the ratio of data to model
reaches zero at energies around 6.3 keV.

We add an absorption edge to our fit, which attenuates the continuum
above $E_{edge}$ with the optical depth $\tau$ =
$\tau_0(E/E_{edge})^{-3}$.  An absorption edge with $E_{edge} = 5.8$
keV and $\tau_0 = 3.5$ significantly improves the fit ($\Delta C = -
15$ with two extra free parameters, see Table 1).  We also tried to
model the absorption feature by a saturated absorption line model.
The model $notch$ of XSPEC was used by fixing the covering fraction at
0.99 to represent a blank absorption trough.
The fitting is slightly better than the edge model with $\Delta C = -
17$ for two extra free parameters, which are located at $E_c = 6.3$
keV, and the line width of $0.9$ keV (both in the observed frame).
The line width can also be taken as the equivalent width of the
absorption line since the absorption is saturated.  We also perform
spectral fits to search for possible other absorption edges/lines at
lower energies. We try different energy entries from 0.5 to 6 keV, but
we find that no further absorption edge/line is statistically required
($\Delta C < 3$).

The X--ray spectrum of CDFS11 was found to be variable with high 
probability ($> 3 \sigma$, Paolillo et al. 2004).
However, due to the limited number of X--ray photons
over 5 keV, and the fact that the absorption feature is optically
thick with black trough, we are unable to study the variability of the
absorption itself, based on our data.
Sometimes, a broad emission line might actually be mimicked by a
strong absorption edge at higher energy (e.g., Reeves, Porquet \& Turner 2004) and
vice versa, especially in X--ray spectra with low S/N or low energy
resolution.  To check if this is the case for CDFS11, we fit the
spectrum with a power law continuum and a broad emission line.  We
tried different central values for the line energy from 4 to 6 keV,
and found that an emission line can only improve the C statistics by
$\Delta C \le - 2$.  This indicates that the absorption feature in
CDFS11 can't be ascribed to the presence of a broad emission line.

\section{Discussion}
\label{lineorigin}

We discuss here the possible origin of the significant broad
absorption feature we detected in the X--ray spectrum of CDFS11 with
the 1Ms $Chandra$ ACIS exposure.  The feature locates at energies $>
20$ keV in the rest frame, with a confidence level of 99.98\%
according to F--test.  We note that the confidence level given by
F--test for an absorption feature might not be accurate
(Protassov et al. 2002).  Here we
re--estimate the confidence level of the absorption feature via a
simple approach.  In the energy range $5.8 - 6.8$ keV where the
absorption feature is located, we detected a total of 11 photons
(source + background), while the continuum model plus background
predict 32 photons.  The cumulative Poisson probability for such a
deviation is $\sim 1.7\times 10^{-5}$.  Since the absorption feature
was examined over the whole spectral band, which is $0.5 - 9.0$ keV,
the probability of detecting such a broad spurious absorption feature
randomly in the spectrum is approximately $1.7 \times 10^{-5}
/(9.0-0.5) = 2.0 \times 10^{-4}$.  This is consistent with the
confidence level obtained with the F--test.

We also perform extensive Monte Carlo simulations to check the
confidence level. First, we simulate
$10^4$ artificial spectra based on the best--fitting continuum model.
We then search for spurious absorption lines in the artificial spectra
by adding a notch component.  All the three parameters of the notch
model (central energy, width, and covering factor) are thawed.  We did
not fix the covering factor at 0.99 during the simulation because
shallower absorption line (i.e., with lower covering factor) can reach
comparable significance level at lower energy due to the higher S/N.
To ensure an efficient search over the whole band, we perform the
search in narrow energy bins (such as $0.5 - 1.5$ keV, $1.5 - 2.5$
keV, etc) with width comparable to that of the observed absorption
feature, and count the total number of spectra in which we detected
spurious absorption line with $\Delta C < -17$.  We found only 6
spectra with statistically significant broad (energy width $> 0.5$
keV) spurious absorption lines.  This indicates a confidence level of
$99.94$\% for the broad absorption feature we detected. Note also
that, while fitting the real spectra, we use only 2 free parameters of
the notch model, while in the simulation, we use 3. We remark that the
confidence level would be higher ($99.98$\%) if we searched for spurious
features with $\Delta C < -19.6$ for 3 free parameters, instead of
$\Delta C < -17$ for 2.

We note that there is also weak evidence of the absorption feature in
the 370 ks XMM spectrum (Streblyanska et al. 2004) of CDFS11. The XMM
spectrum is significantly steeper ($\Gamma = 2.3$) and it was obtained
about one year later.  The statistical quality of the XMM spectrum
around 6 keV is much lower due to the high background noise and the
steeper continuum.  By fitting the Chandra and XMM spectra
simultaneously, we found an even higher confidence level of 99.997\%
for the absorption feature, based on F--test. However, due to the very
limited statistical quality, we are unable to use the XMM spectrum
independently to constrain the nature and possible time variability of
the absorption feature.

We conclude that the confidence level of the absorption line we
detected in CDFS11 is $> 99.98$\%.  Even considering the number of all
the CDFS spectra we examined for interesting features ($\sim 30$ CDFS
spectra have X--ray photons $> 500$, whose quality is high enough to
enable the search for broad line features), the absorption feature is
still significant with a confidence level of $99.4$\%.

\subsection{Instrumental Artifact?}
Could such an unusual X--ray feature be due to any instrumental
artifact or to some aspect of our analysis?  We first consider
improper background subtraction as a possible origin of the absorption
feature.  We check that we obtain consistent results when using
background spectra extracted from different regions of the detector.
We also note that the absorption feature is already significant in the
spectrum {\it without} background subtraction.  Finally, we do not see
any bump at $\sim 6$ keV in the background spectrum either, suggesting
that the significance level of the feature is not magnified by
background subtraction.  Therefore, we conclude that the absorption
feature is not related to uncertainties in the background subtraction.

Could the absorption feature be an artifact of calibration uncertainty
in the ACIS instrumental response function? As far as we know, no such
an artifact (i.e., an absorption feature at $\sim 6.3$ keV) has been
reported.  We examined the spectra of nearby X--ray sources, but we
found no evidence of absorption feature at $\sim 6.3$ keV among them.
Tozzi et al. (2005) presented the X--ray spectra of CDFS sources using
updated $Chandra$ calibration files (with CIAO3.0.1 and CALDB2.26
instead of CIAO2.0.1 and CALDB2.0 adopted in Giacconi et al. 2002 and
in this letter).  We repeated the fitting using the updated spectrum
and obtained consistent results for both the continuum and the
absorption feature.  This confirms that the absorption feature is not
due to improper data calibration.  Note that Wang et al. (2003)
reported a puzzling strong emission line at $\sim 6.3$ keV in the
observed frame in CXO CDFS J033225.3-274219 (CDFS46). The two unusual
features (the emission feature in CDFS46, and the absorption feature
in CDFS11 reported in this letter) located at similar energies in the
observing frame, could suggest a common origin related to instrumental
effects. However, there are strong evidences against this occurrence:
since the total exposure was composed of eleven individual
observations with different roll angles, photons from each source fall
into different positions on different chips during different
observations.  It is unlikely that an unknown instrumental artifact
(if there is any) would affect photons from all eleven
exposures. Furthermore, the two sources are around 9\arcmin\ apart,
and fall into different chips during the observations. We conclude
that the similar energy of the two features is just a coincidence.

\subsection{The nature of the absorber}
Among the heavy elements, iron is the only one that can produce such a
strong absorption feature at high energies. However, the rest--frame
energy of the absorption ($> 20$ keV) is too high for static iron in
the rest frame, since the highest energy transition is the K--shell
edge of H--like ion at 9.28 keV. Therefor we consider two
possibilities: either the absorber locates at much lower redshift
along the line of sight, or in an relativistic outflow intrinsic to
the quasar.

The continuum fitting yields a marginal intrinsic absorption
with $N_H \sim 10 ^{21}$ cm$^{-2}$, which is too small to
account for the strong absorption feature at $\sim 6.3$ keV.  We
can't locate other absorption features either in the spectrum at lower
energy (due to relatively lower ionized ions, such as the OVIII
absorption edge at 0.87 keV).  This suggests that the absorber has to
be extremely ionized, with almost all abundant elements fully ionized,
and the absorption is dominated by H or He--like iron atoms.  
Adopting the photon ionization absorption model $absori$ (Magdziarz \&
Zdziarski 1995) to fit the spectrum, we obtain an ionization
parameter\footnote{$\xi$\ = $L/nR^2$, where $L$ is the integrated
incident luminosity between 5 eV and 300 keV, $n$ is the density of
the material and $R$ is the distance of the material from the
illuminating source (Done et al. 1992).} $\xi > 9000$ and $N_H > 3
\times 10^{24}$ cm$^{-2}$ (Table 1). 
Note the upper limits of $\xi$\ and $N_H$ are poorly constrained since
the two parameters are degenerate.
Assuming $\xi = 10000$, we obtain $N_H = 5 \times 10^{24}$ cm$^{-2}$.
We assume a temperature of $10^6$ K and solar abundance for the
absorber.  We note that an intrinsic absorption at the level of $N_H =
5 \times 10^{24}$ cm$^{-2}$ (well within the Compton--thick regime)
would strongly attenuate the continuum photons.
However, if the iron in the absorber is
a factor of 10 overabundant with respect to the solar value, we obtain
$N_H \sim 5 \times 10^{23}$ cm$^{-2}$.

The absorption feature can also be fitted by a broad saturated
absorption line. The most likely responsible lines are 6.70 Fe$_{XXV}$
K$\alpha$ (6.70 keV) and/or Fe$_{XXVI}$ K$\alpha$ (6.97 keV).
Assuming an average depth of the absorption line $\tau = 2$, and a
factor of 10 overabundant for iron, we obtain $N_H \sim 5 \times
10^{23}$ cm$^{-2}$.  We point out that for such an absorber, we also
expect Fe$_{XXV}$ K$\beta$ (7.85 keV), Fe$_{XXVI}$ K$\beta$ (8.26
keV) absorption lines, and strong Fe$_{XXV}$ (8.83 keV), Fe$_{XXVI}$
(9.28 keV) absorption edges.  Given the few counts we measure above 7
keV, we can't check the existence of these features.  Future X--ray
missions with higher sensitivity can help us to unveil the nature of the
absorber by showing us the spectral properties at energies $> 7$ keV
and the absorption profile with higher S/N.

If the absorption is due to foreground absorber, the absorber has to
be located at a much lower redshift $\sim 0.5$ (if due to H or He-like
iron absorption edges) or $\sim 0.1$ (if due to H or He-like iron
resonant absorption line). The absorber must be extremely ionized with
$N_H > 5 \times 10^{23}$ cm$^{-2}$.  Note the foreground absorber is
unlikely photo--ionized, otherwise we should have seen the extra
photon ionization source, thus it has to be extremely hot with a
temperature of $\sim 10^8$ K to reach the required ionization
stage. Such a high temperature and $N_H$ is very unusual for
intervening systems.  The only possible candidate is the hot
intra--cluster medium found in clusters of galaxies (see Rosati,
Borgani \& Norman 2002 and references therein).  However, for a cluster with $T
= 10^8$ K, we expect an X--ray luminosity of $\sim 3 \times 10^{45}$
erg s$^{-1}$ (based on the $L_X - T$ relation, e.g.  Lumb et
al. 2004).  If located at z $\sim 0.5$, such a cluster would be around
300 times brighter than CDFS11, and its emission would be obviously
prominent in the {\sl Chandra} image.  We conclude that the X--ray
absorption feature we detected in CDFS11 can't be due to an
intervening systems along the line of sight.

Finally, we consider a relativistic outflow intrinsic to the quasar,
with speed of $\sim 0.7 c$ (Doppler Factor $DF = 2.3$ for absorption
edge) or $\sim 0.83 c$ ($DF = 3.2$ for absorption line), to produce
the observed blueshift of the absorption feature.
Note that similar blueshifted features have been reported in the
X--ray spectra of 3 AGNs: a highly blueshifted O$_{VII}$ emission line 
in PKS 0637-75 with $DF \sim 2.7 - 2.8$ (Yaqoob et al. 1998); a blueshifted 
Fe--K emission line ($DF \sim 2.4- 2.6$) in QSO PKS 2149-306 
(Yaqoob et al.  1999); and a strong blueshifted Fe-K emission line ($DF \sim
2.3 - 2.5$) in CXO CDFS J033225.3-274219 (Wang et al. 2003).  
The good agreements of the blueshift factors from different sources
strongly suggest a close origin of the relativistic outflow, and
strengthen the statistical significance of these features.

Taking log$\xi = 4$ and $N_H = 5 \times 10^{23}$ cm$^{-2}$, we estimate 
the distance of the absorber from the central source $R$ by assuming the thickness
of the absorber $\Delta R \sim 0.1 R$. We obtain $R \sim 6 \times
10^{16}$ cm.  Following Reeves, O'Brien \& Ward (2003), we estimate the
rate of relativistic outflow.  The outflow rate ($\dot{M}_{\rm out}$)
can be calculated as $\dot{M}_{\rm out}=\Omega v_{\rm out} m_{\rm p}
L_{\rm X}/\xi$.  Assuming that the outflow subtends a solid angle of
0.1~steradian, we obtain a outflow rate of $\sim 10^{27}$ g s$^{-1}$
or $\sim 10 M_{\odot}$~year$^{-1}$, similar to that of the outflow in
PDS 456 ($v_{\rm out} \sim 0.17 c$, Reeves et al. 2003). However, because 
of the extremely high outflow speed in this case, the kinetic energy carried by
the relativistic material outflow would be $\sim 20
M_{\odot}$~year$^{-1}$, and then requires an accretion rate of at
least 200 $M_{\odot}$~year$^{-1}$ with an assumed energy production
efficiency $\epsilon = 0.1$ which entirely goes into the outflow.
Such an accretion rate is far beyond the Eddington limit even for a
supermassive black hole of $10^8M_{\odot}$ ($\sim 3
M_{\odot}$~year$^{-1}$).  To keep the rate of kinetic energy outflow
lower than that released by Eddington accretion, a more reasonable
outflow rate would be at least 100 times smaller ($< 0.1
M_{\odot}$~year$^{-1}$), and so does the solid angle ($< 0.001$
steradian). The extremely high velocity and small covering
factor strongly link the physics of the outflow to relativistic jets
in AGN and X-ray binary systems (such as the famous XRB SS 433, see
Migliari, Fender \& M\'{e}ndez 2002). 

We conclude that the X-ray absorption feature in CDFS11 is due to
an ionized jet-like outflow intrinsic to the quasar, with a bulk velocity 
of $\sim$ 0.7 - 0.8 c. Similar outflows have been reported in the
X-ray spectra of several other AGNs. Future X-ray mission with
higher intensity can help us understand the nature of the outflows
by providing X-ray spectra with higher S/N.

\acknowledgments
The work of JW was supported by Chinese NSF through NSF10473009 and the CAS 
"Bai Ren" project at University of Science and Technology of China.
\clearpage
\begin{deluxetable}{llcc}
\tablecaption{Spectral fits to CDFS11}
\tablecolumns{6}
\tablewidth{0pt}
\startdata
\hline
continuum &$\Gamma$ &N$_H$(10$^{22}$cm$^{-2}$) &C/dof\\
	  & 1.7$^{+0.1}_{-0.1}$ &0.1$^{+1.0}_{-0.1}$ & 362/356\\\hline
edge      &E$_{edge}$(keV) &$\tau_0$ & C/dof\\
          &20.9$^{+0.7}_{-0.5}$ &3.5$^{+...B}_{-1.2}$ & 347/354\\\hline
notch$^A$ 	  &E$_c$(keV) &Width(keV) & B/dof\\
          &22.5$^{+0.5}_{-0.3}$ &3.0$^{+0.9}_{-1.1}$ & 345/354\\\hline
absori    &$\xi$    &N$_H$(10$^{22}$cm$^{-2}$) &C/dof\\
	  &$>$9000    &$>$500          &344/353
\enddata
\tablenotetext{A}{The covering factor of the notch model was fixed to 0.99 to 
represent a heavily saturated absorption.} 
\tablenotetext{B}{The upper limit of the absorption depth was poorly constrained.}
\end{deluxetable}

\clearpage

\begin{references}  
\reference{}Alexander, D. M., et al. 2003, AJ, 126, 539
\reference{}Brandt, W. N., et al. 2001, AJ, 122, 2810
\reference{}Cash, W. 1979, ApJ, 228, 939
\reference{}Dickey, L. M., \& Lockman, F. J. 1990, ARA\&A. 28, 215. 
\reference{}Done, C., Mulchaey, J. S., Mushotzky, R. F., \& Arnaud, K. A. 1992, ApJ, 395, 275
\reference{}Giacconi, R., Gurksy, H., Paolini, F.R., \& Rossi, B.B., 1962,
  Phys. Rev. Letters, 9, 439
\reference{}Giacconi, R., et al. 2002, ApJS, 139, 369
\reference{}Hornschemeier, A. E., et al. 2001, ApJ, 551, 742
\reference{}Lumb, D. H. et al. 2004, A\&A, 420, 853
\reference{}Magdziarz, P., \& Zdziarski, A. A. 1995, MNRAS, 273, 837
\reference{}Migliari, S., Fender, R., \& M\'{e}ndez, M. 2002, Science, 297, 1673
\reference{}Nousek, J. A., \& Shue, D. R. 1989, ApJ, 342, 1207
\reference{}Paolillo, M. et al. 2004, ApJ, 611, 93
\reference{}Protassov, R.; van Dyk, D. A., Connors, A., Kashyap, V. L., \& Siemiginowska, A, 2002, ApJ, 571, 545
\reference{}Reeves, J. N., O'Brien, P. T., \& Ward, M. J. 2003, ApJ, 593, L65
\reference{}Reeves, J. N., Porquet, D., \& Turner, T. J. 2004, ApJ, 615, 150
\reference{}Rosati, P., et al. 2002, ApJ, 566, 667
\reference{}Rosati, P., Borgani, S., \& Norman, C. 2002, ARA\&A, 40, 539
\reference{}Streblyanska, A., Bergeron, J., Brunner, H., Finoguenov, A., Hasinger, G., Mainieri. V. 2004, $Nuc.~Phys.~B~(Proc.~Supp.)$, 132, 232
\reference{}Szokoly, G. P., et al. 2004, ApJS, 155, 271
\reference{}Tozzi, P, et al. 2005, A\&A submitted
\reference{}Wang, J. X. et al. 2003, ApJ, 590, L87
\reference{}Yaqoob, T., George, I.M., Turner, T.J., Nandra, K., Ptak, A., \&
  Serlemitsos, P.J. 1998, 505, L87
\reference{}Yaqoob, T., George, I.M., Nandra, K., Turner, T.J., Zobair, S., \&
  Serlemitsos, P.J. 1999, ApJ, 525, L9
\end{references}
\end{document}